\pdfoutput=1

\documentclass[11pt]{article}

\usepackage[review]{acl}

\usepackage{times}
\usepackage{latexsym}
\usepackage{url}
\usepackage[T1]{fontenc}

\usepackage[utf8]{inputenc}

\usepackage{microtype}

\usepackage{inconsolata}

\usepackage{graphicx}

\usepackage{float}
\usepackage{hyperref}
\usepackage{amsthm,amsmath,amssymb}
\usepackage{mathrsfs}
\usepackage{enumitem}       

\usepackage{afterpage}

\usepackage{graphicx}

\title{Continuous Speech Tokenizer in Text To Speech}

\author{
 \textbf{Yixing Li\textsuperscript{1}},
 \textbf{Ruobing Xie\textsuperscript{2}}*,
 \textbf{Xingwu Sun\textsuperscript{2,3}},
 \textbf{Yu Cheng\textsuperscript{1}*},
 \textbf{Zhanhui Kang\textsuperscript{2}}
\\
 \textsuperscript{1}The Chinese University of Hong Kong~~ 
 \\
 \textsuperscript{2}Machine Learning Platform Department, Tencent
 \\
 \textsuperscript{3}University of Macau
 \\
\texttt{lyxing0@sjtu.edu.cn}~~~
\texttt{xrbsnowing@163.com}
\\
\texttt{sammsun@tencent.com}~~~  
\texttt{chengyu@cse.cuhk.edu.hk}
}

\begin{document}
\maketitle
\begin{abstract}
The fusion of speech and language in the era of large language models has garnered significant attention. Discrete speech token is often utilized in text-to-speech tasks for speech compression and portability, which is convenient for joint training with text and have good compression efficiency. However, we found that the discrete speech tokenizer still suffers from information loss. Therefore, we propose a simple yet effective continuous speech tokenizer named \textbf{Cont-SPT},  and a text-to-speech model based on continuous speech tokens. Our results show that the speech language model based on the continuous speech tokenizer has better continuity and higher estimated Mean Opinion Scores (MoS). This enhancement is attributed to better information preservation rate of the continuous speech tokenizer across both low and high frequencies in the frequency domain. The code and resources for 
\textbf{Cont-SPT} can be found in \href{https://github.com/Yixing-Li/Continuous-Speech-Tokenizer}{https://github.com/Yixing-Li/Continuous-Speech-Tokenizer}.

\end{abstract}

\section{Introduction}

\let\thefootnote\relax\footnotetext{* Corresponding author.}

Autoregressive modeling is a common method for processing language sequences and is effective in token prediction \cite{vaswani2017attention, radford2018improving, yang2024harnessing}. The success of large language models continues to inspire and develop this field \cite{Tay2020EfficientTA, dao2022flashattention, zhuang2023survey}. Transformer-based models are becoming more and more prominent due to the interest in speech language models utilized to understand and generate speech \cite{audiolm, VALLE,  Audiopalm, Speechx}. These models are designed to solve a range of speech-related tasks such as Text-to-Speech (TTS) \cite{tts_overview} and Automatic-Speech-Recognition (ASR) \cite{asr_overview}. A significant difference between the text and speech domains is the discrete nature of text tokens versus the continuous nature of speech, whereas a commonality of language models is the utilization of discrete speech representations \cite{wang2023viola, Uniaudio, tts_overview}. %
The current discrete speech representation can be divided into two types: tokenizer-based compressed tokens such as VALL-E \cite{VALLE}, MusicLM \cite{Musiclm}, CosyVoice \cite{du2024cosyvoice}, and semantic tokens such as AudioLM \cite{audiolm}. The tokenizer-based model utilizes residual vector quantization (RVQ) and hierarchical quantizers to compress speech tokens, and semantic tokens are usually based on self-supervised pre-trained models.
Many works have been studying the use of speech tokens, but the best speech representation remains unresolved \cite{encodec, yang2023hifi, speechtokenizer}.


We observe that discrete speech tokenizers have potential information loss in text-to-speech tasks, which is reflected in different degrees of loss in low-frequency and high-frequency components. 
Therefore, we utilized a speech encoder that models \textbf{continuous speech vectors} for TTS. 
Based on this, we conducted experiments to demonstrate the advantages of continuous speech tokens over discrete versions. Continuous speech tokens have better information retention than discrete tokens in all frequency bands and better robustness to the sampling rate in text-to-speech tasks. 
Specifically, our contributions include:
\begin{enumerate}[leftmargin=*]
    \item We proposed a continuous speech tokenizer based text-to-speech model and provided a comprehensive training framework.
    \item We conducted experiments to demonstrate the effectiveness of our approach in the text-to-speech task and showed the significance of the training pipeline.
    \item We verified the advantages of the continuous speech tokenizer over the discrete version, including the retention of high-frequency information and the robustness of the sampling rate.
\end{enumerate}




\begin{figure*}[!t]
  \includegraphics[width=0.98\linewidth]{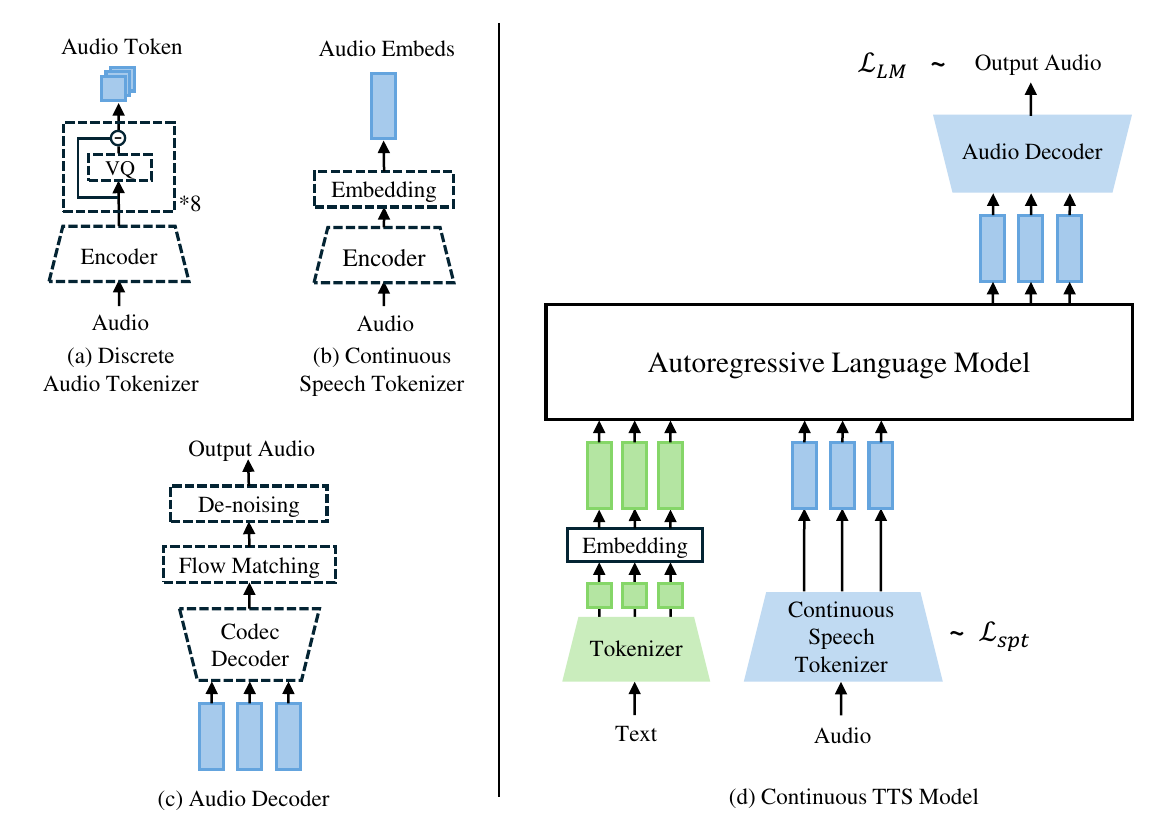} \hfill
  \caption {Continuous Text-to-speech Model. (a) Traditional audio tokenizer structure, using preset or trainable code table for RVQ. (b) \textbf{Continuous speech tokenizer}, replace RVQ with an embedding module to generate continuous speech token representations. (c) Audio decoder, consisting of codec decoder, flow-matching and de-noising modules. (d) Overall continuous TTS model structure.}
  \label{fig:model}
\end{figure*}

\section{Method}

\subsection{Continuous Speech Tokenizer}

Most contemporary speech language models utilize the RVQ quantizer \cite{encodec, VALLE, Musiclm} as a basic building block to accomplish the process of speech discretization. As illustrated in Figure \ref{fig:model}a, the quantizer usually contains multiple codebooks that convert speech features into discrete tokens. The quantization process usually involves distance grounding based on the content within the codebook, which suffers from great information loss. 


We propose to adopt a \textbf{continuous speech tokenizer (Cont-SPT)} for text-to-speech tasks. After the input audio is resampled at 24 kHz, it is passed through the encoder to obtain the speech feature vector. In the discrete version, this step is typically followed by RVQ to generate discrete tokens. In contrast, our continuous speech tokenizer retains the audio resampling step and integrates the encoder layer to obtain continuous speech tokens, and then utilizes the output continuous speech tokens directly as the input for the language model.

Figure \ref{fig:model}b shows the continuous speech tokenizer. The decoder corresponds to the encoder, which is utilized to decode the output of the autoregressive language model. The two can be pre-trained together in a VAE-like form, and our experiments show that this is beneficial for TTS tasks.

\subsection{Continuous Speech Tokenizer Based TTS}

We propose a TTS model based on the continuous speech tokenizer and regards the TTS task as an autoregressive token generation task. Different from predicting discrete speech codes, our method predicts continuous speech tokens. Given the input speech $\mathbb{A}$ and text $\mathbb{T}$, we calculate the continuous speech token and text embedding respectively,
\begin{equation}
    \mathcal{A} = \text{Cont-SpeechTokenizer}\left(\mathbb{A}\right).
\end{equation}
where $\mathcal{A}$ denotes for the continuous speech token. Text processing includes text tokenizer and language model embedding layer,
\begin{equation}
   x = \text{Text Tokenizer}\left(\mathbb{T}\right), \\
\mathcal{T} = \text{Embeds}\left(x\right).
\end{equation}
The speech token and text embedding are concatenated together as the input for the language model during autoregressive generation. After generating the predicted continuous speech token, we utilize the audio decoding layer and flow matching to convert the audio.

\subsection{Flow Matching}

We utilize the optimal-transport conditional flow matching (OT-CFM) following the codec decoder to 
generate a speech mel-spectrogram based on the generated continuous speech features.
By leveraging the idea of optimal transport, CFM can be constructed to generate an ordinary differential equation (ODE) with simple vector fields.

Moreover, de-noising module is added for final output adjustment, consisting of frequency limiting and off-the-shelf denoising model.

\subsection{Training Objective}

\subsubsection{Speech Tokenizer Reconstruction Loss}\label{train1}

For the pre-training of the speech tokenizer, we use a training method similar to VAE, connecting the continuous speech tokenizer with the speech decoder for overall training. Given the training audio $\mathbb{A}$ and the corresponding text label $\mathbb{Y}$, the calculation process is as follows,

\begin{equation}
    \widehat{\mathbb{A}} = \text{AD}\left(\text{Cont-SpeechTokenizer}\left(\mathbb{A}\right)\right).
\end{equation}

where AD refers to AudioDecoder. And the estimated audio-to-text is calculated by the ASR decoder:

\begin{equation}
    \widehat{\mathbb{Y}} = \text{ASR-Decoder}\left( \widehat{\mathbb{A}}\right).
\end{equation}

The reconstruction loss contains two parts, the similarity loss between the regenerated audio $\widehat{\mathbb{A}}$  and the original signal $\mathbb{A}$, and the CTC loss between the decoded ASR results $\widehat{\mathbb{Y}}$ and the text label $\mathbb{Y}$ .
\begin{equation}
\begin{aligned}
    \mathcal{L}_{spt} &= \mathcal{L}_{rec} + \mathcal{L}_{ASR} \\
    &=\text{SIM}\left( \mathbb{A}, \widehat{\mathbb{A}} \right) + \text{CTC}\left( \mathbb{Y}, \widehat{\mathbb{Y}} \right). 
\end{aligned}
\end{equation}

\subsubsection{Language Modeling Loss}\label{train2}
In the language model training part, we also activate the speech continuous tokenizer as a training state and train it together with the language model, while these two components are updated with different learning rates. Given the model output $\mathcal{O}$ and the speech label $\mathbb{M}$, the speech label is first encoded into continuous speech tokens, and the loss function is calculated by the distance between the continuous speech token of the label and the model output.

\begin{equation}
    \mathcal{M} = \text{Cont-SpeechTokenizer}\left( \mathbb{M}\right).
\end{equation}
\begin{equation}
    \mathcal{L}_{LM} = \text{MSE}\left( \mathcal{O}, \mathcal{M}\right).
\end{equation}

\subsubsection{Training Pipeline}

The training consists of two steps, the pre-training of the tokenizer and the joint training of the entire model. First, the tokenizer encoder is connected to the decoder and pre-trained as described in \ref{train1}. In the second step, the decoder is frozen, and the entire model is jointly trained as outlined in \ref{train2}. During this step, the learning rate of the tokenizer is set to 0.05 of that of the language model.

\section{Experiments}

\subsection{Experiment Setup}

\noindent
\textbf{Datasets.}  
For training, we utilize the  LibriSpeech \cite{Librispeech} dataset, which contains a corpus of approximately 1,000 hours of English speech. We use the dev-clean subset for validation and the test-clean subset for evaluation.

\noindent
\textbf{Model Settings.}
Following VALL-E \cite{VALLE}, we construct the continuous speech tokenizer based TTS model, while the number of model blocks are set to the default values. 

\noindent
\textbf{Baselines.}
For the speech tokenizer experiments, we use the commonly used discrete speech tokenizer Encoder as the baseline. For the TTS model experiment, we use the following models as the baseline: VALL-E \cite{VALLE}, MELLE \cite{melle}. 

\noindent
\textbf{Evaluation Metrics.}
For the speech tokenizer, we utilize an estimated transfer function to show the ability of the speech tokenizer in preserving information. For the TTS model, we employ the following metrics: (1) WER (Word Error Rate), (2) SIM (Speaker Similarity), (3) EMoS (Estimated Mean Opinion Score), (4) CLVP Score, (5) STOI, (6) Noisiness, Continuity, Loudness Quality, and Naturalness. Details of the metrics are shown in the Appendix \ref{app:metric}.

\subsection{TTS Evaluation}

\begin{table}[!t]
  \centering
  \begin{tabular}{ccc}
    \hline
    \textbf{Model $\backslash$  Metrics} & \textbf{WER $\downarrow$}  & \textbf{SIM $\uparrow$} \\
    \hline
    Origin     & 2.61  &     0.86       \\
    VALL-E     & 12.73  &    0.53        \\
    MELLE      & 7.26  &     0.67        \\
    Ours       & \textbf{6.59}  &   \textbf{0.73}          \\
    \hline
  \end{tabular}
  \caption{Evaluation of WER ($\%$) and SIM on the LibriSpeech. }
  \label{tab:metrics1}
\end{table}

\begin{table}[!t]
  \centering
  \begin{tabular}{ccc}
    \hline
    \textbf{Metrics} & \textbf{VALL-E}  & \textbf{Cont-SPT} \\
    \hline
    EMoS $\uparrow$     & 0.83  &\textbf{ 1.32}           \\
    CLVP $\downarrow$     & 3.52  & \textbf{2.94}            \\
    STOI $\uparrow$    & 0.20  & \textbf{0.42}            \\
    Noisiness Quality $\uparrow$      & 1.27   & \textbf{1.82}            \\
    Continuity Quality $\uparrow$     & 1.80  & \textbf{3.61}           \\
    Loudness Quality $\uparrow$     & 1.56  & \textbf{1.65}           \\
    Naturalness $\uparrow$   & 1.31  & \textbf{1.45}           \\
    \hline
  \end{tabular}
  \caption{Evaluation of speech generation quality. Cont-SPT that based on continuous speech tokens achieved higher speech quality than baseline model based on discrete tokens. }
  \label{tab:metrics2}
\end{table}

In this section we evaluate and compare the results of the continuous speech tokenizer based TTS model.
Our text-to-speech model based on continuous speech tokens achieves better performance in various aspects and shows its value in TTS tasks.

Table \ref{tab:metrics1} shows the comparison between our method and the baseline, and our method outperforms the baseline in most cases. 
More detailed evaluation indicators shown in Table \ref{tab:metrics2} elucidate the specific advantages of our method. In the evaluation of relative indicators including CLVP Score, STOI, and absolute indicators such as noisiness quality and continuity quality, the model based on continuous speech tokenizer exceeds the baseline. The ablation study is shown in Appendix \ref{app:ablation}.

\subsection{Analysis on Different Frequencies}\label{sec:eval freq}

\begin{table}[!t]
  \centering
  \begin{tabular}{cccc}
    \hline
    \textbf{SPT $\setminus$  Freq} & \textbf{2k}  & \textbf{5k}& \textbf{8k} \\
    \hline
    Discrete      & 0.95   & 0.78 & 0.34            \\
    Continuous      & 0.94  &0.81 & 0.55             \\
    \hline
  \end{tabular}
  \caption{The retention rate of audio after the tokenizer at different frequencies. The continuous speech tokenizer has high information retention in all frequency bands, especially in the high frequency part. }
  \label{tab:metrics2_freq}
\end{table}

We are interested in how continuous speech tokenizers can help TTS models improve their performance. As mentioned above, the number of tokens in the codebook of discrete tokenizers is limited, which constrains the space that discrete tokenizers can represent to be limited and increases the information loss. In this section, we demonstrate the advantages of continuous speech tokenizers in retaining information by studying the transfer function of speech tokenizers.


Table \ref{tab:metrics2_freq} presents the estimated transfer functions of the continuous and discrete speech tokenizers across various frequency ranges. It can be seen that in the high-frequency range, the transmission effect of the discrete speech tokenizer drops sharply, while the continuous speech tokenizer maintains a good effect in this range. In fact, the continuous speech tokenizer has a great information retention ability in the entire frequency range.

\subsection{Sampling Rate and Window Length Robustness}


\begin{figure}[!t]
  \includegraphics[width=0.95\columnwidth]{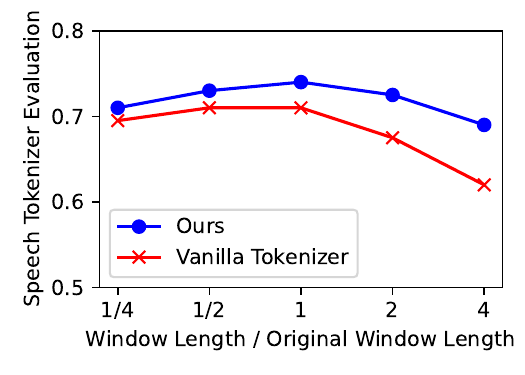}
  \caption{Evaluation of modifying window length and sampling rate utilizing the average attention rate after the tokenizer.}
  \label{fig:WL}
\end{figure}

We further study the robustness to different sampling rates. Due to the presence of narrowband data, robustness to sampling rate is also worth noting. 
To ensure the consistency of input dimensions, we modify the sampling rate and window length synchronously. As shown in Figure \ref{fig:WL}, the continuous speech tokenizer is more robust than the discrete tokenizer, especially when the window length ratio is larger. This is due to the better information preservation of continuous speech tokenizer, and this conclusion is consistent with Section \ref{sec:eval freq}.

\section{Conclusion}

In this paper, we propose a method based on continuous speech tokenizer in TTS and demonstrate its effectiveness through experiments. In addition, we conduct a series of experiments on continuous and discrete speech tokenizers to illustrate the advantages of continuous tokenizer in terms of information retention and robustness. Finally, the TTS model based on continuous speech token proposed in this paper provides a complete framework and a foundation for subsequent research.

\section*{Limitations}

Although we proposed a novel continuous speech token based approach in the TTS field and compared it with discrete methods, there are limitations for future studies. Our work focuses on TTS tasks as well as joint training of speech and text, and has not yet been verified and evaluated on the Multimodal Large Language Model (MLLM) \cite{mmllm}. The addition of multiple modalities brings challenges. The continuous speech tokenizer proposed in our work could improve the information carrying capacity, but it may also bring difficulties of training. Future research will continue to explore these unexplored aspects.

\section*{Acknowledgments}

This work is supported by the Young Elite Scientists Sponsorship Program by CAST (2023QNRC001).


\bibliography{main}

\begin{thebibliography}{31}
\providecommand{\natexlab}[1]{#1}

\bibitem[{SPE()}]{SPEECHMETRICS}

\newblock A wrapper around speech quality metrics mosnet, bsseval, stoi, pesq, srmr, sisdr.
\newblock [Online].
\newblock \url{https://github.com/aliutkus/speechmetrics.git}.

\bibitem[{Agostinelli et~al.(2023)Agostinelli, Denk, Borsos, Engel, Verzetti, Caillon, Huang, Jansen, Roberts, Tagliasacchi et~al.}]{Musiclm}
Andrea Agostinelli, Timo~I Denk, Zal{\'a}n Borsos, Jesse Engel, Mauro Verzetti, Antoine Caillon, Qingqing Huang, Aren Jansen, Adam Roberts, Marco Tagliasacchi, et~al. 2023.
\newblock Musiclm: Generating music from text.
\newblock \emph{arXiv preprint arXiv:2301.11325}.

\bibitem[{Betker(2022)}]{TTS-scores}
J~ames Betker. 2022.
\newblock \href {https://github.com/neonbjb/tts-scores} {{TTS-scores}}.

\bibitem[{Borsos et~al.(2023)Borsos, Marinier, Vincent, Kharitonov, Pietquin, Sharifi, Roblek, Teboul, Grangier, Tagliasacchi et~al.}]{audiolm}
Zal{\'a}n Borsos, Rapha{\"e}l Marinier, Damien Vincent, Eugene Kharitonov, Olivier Pietquin, Matt Sharifi, Dominik Roblek, Olivier Teboul, David Grangier, Marco Tagliasacchi, et~al. 2023.
\newblock Audiolm: a language modeling approach to audio generation.
\newblock \emph{IEEE/ACM transactions on audio, speech, and language processing}, 31:2523--2533.

\bibitem[{Chen et~al.(2022)Chen, Wang, Chen, Wu, Liu, Chen, Li, Kanda, Yoshioka, Xiao et~al.}]{chen2022wavlm}
Sanyuan Chen, Chengyi Wang, Zhengyang Chen, Yu~Wu, Shujie Liu, Zhuo Chen, Jinyu Li, Naoyuki Kanda, Takuya Yoshioka, Xiong Xiao, et~al. 2022.
\newblock Wavlm: Large-scale self-supervised pre-training for full stack speech processing.
\newblock \emph{IEEE Journal of Selected Topics in Signal Processing}, 16(6):1505--1518.

\bibitem[{Dao et~al.(2022)Dao, Fu, Ermon, Rudra, and R{\'e}}]{dao2022flashattention}
Tri Dao, Dan Fu, Stefano Ermon, Atri Rudra, and Christopher R{\'e}. 2022.
\newblock Flashattention: Fast and memory-efficient exact attention with io-awareness.
\newblock \emph{Advances in Neural Information Processing Systems}, 35:16344--16359.

\bibitem[{D{\'e}fossez et~al.(2022)D{\'e}fossez, Copet, Synnaeve, and Adi}]{encodec}
Alexandre D{\'e}fossez, Jade Copet, Gabriel Synnaeve, and Yossi Adi. 2022.
\newblock High fidelity neural audio compression.
\newblock \emph{arXiv preprint arXiv:2210.13438}.

\bibitem[{Du et~al.(2024)Du, Chen, Zhang, Hu, Lu, Yang, Hu, Zheng, Gu, Ma et~al.}]{du2024cosyvoice}
Zhihao Du, Qian Chen, Shiliang Zhang, Kai Hu, Heng Lu, Yexin Yang, Hangrui Hu, Siqi Zheng, Yue Gu, Ziyang Ma, et~al. 2024.
\newblock Cosyvoice: A scalable multilingual zero-shot text-to-speech synthesizer based on supervised semantic tokens.
\newblock \emph{arXiv preprint arXiv:2407.05407}.

\bibitem[{Ivan(2024)}]{nisqa_github}
Beskrovnyi Ivan. 2024.
\newblock nisqa-s.
\newblock https://github.com/deepvk/nisqa-s.

\bibitem[{Kheddar et~al.(2024)Kheddar, Hemis, and Himeur}]{asr_overview}
Hamza Kheddar, Mustapha Hemis, and Yassine Himeur. 2024.
\newblock Automatic speech recognition using advanced deep learning approaches: A survey.
\newblock \emph{Information Fusion}, page 102422.

\bibitem[{Lo et~al.(2019)Lo, Fu, Huang, Wang, Yamagishi, Tsao, and Wang}]{EMOS}
Chen-Chou Lo, Szu-Wei Fu, Wen-Chin Huang, Xin Wang, Junichi Yamagishi, Yu~Tsao, and Hsin-Min Wang. 2019.
\newblock Mosnet: Deep learning based objective assessment for voice conversion.
\newblock \emph{arXiv preprint arXiv:1904.08352}.

\bibitem[{Meng et~al.(2024)Meng, Zhou, Liu, Chen, Han, Hu, Liu, Li, Zhao, Wu et~al.}]{melle}
Lingwei Meng, Long Zhou, Shujie Liu, Sanyuan Chen, Bing Han, Shujie Hu, Yanqing Liu, Jinyu Li, Sheng Zhao, Xixin Wu, et~al. 2024.
\newblock Autoregressive speech synthesis without vector quantization.
\newblock \emph{arXiv preprint arXiv:2407.08551}.

\bibitem[{Mittag et~al.(2021)Mittag, Naderi, Chehadi, and Möller}]{NISQA}
Gabriel Mittag, Babak Naderi, Assmaa Chehadi, and Sebastian Möller. 2021.
\newblock \href {https://doi.org/10.21437/interspeech.2021-299} {Nisqa: A deep cnn-self-attention model for multidimensional speech quality prediction with crowdsourced datasets}.
\newblock \emph{Interspeech 2021}.

\bibitem[{Panayotov et~al.(2015)Panayotov, Chen, Povey, and Khudanpur}]{Librispeech}
Vassil Panayotov, Guoguo Chen, Daniel Povey, and Sanjeev Khudanpur. 2015.
\newblock Librispeech: an asr corpus based on public domain audio books.
\newblock In \emph{2015 IEEE international conference on acoustics, speech and signal processing (ICASSP)}, pages 5206--5210. IEEE.

\bibitem[{Radford(2018)}]{radford2018improving}
Alec Radford. 2018.
\newblock Improving language understanding by generative pre-training.

\bibitem[{Radford et~al.(2023)Radford, Kim, Xu, Brockman, McLeavey, and Sutskever}]{Whisper}
Alec Radford, Jong~Wook Kim, Tao Xu, Greg Brockman, Christine McLeavey, and Ilya Sutskever. 2023.
\newblock Robust speech recognition via large-scale weak supervision.
\newblock In \emph{International conference on machine learning}, pages 28492--28518. PMLR.

\bibitem[{Rubenstein et~al.(2023)Rubenstein, Asawaroengchai, Nguyen, Bapna, Borsos, Quitry, Chen, Badawy, Han, Kharitonov et~al.}]{Audiopalm}
Paul~K Rubenstein, Chulayuth Asawaroengchai, Duc~Dung Nguyen, Ankur Bapna, Zal{\'a}n Borsos, F{\'e}lix de~Chaumont Quitry, Peter Chen, Dalia~El Badawy, Wei Han, Eugene Kharitonov, et~al. 2023.
\newblock Audiopalm: A large language model that can speak and listen.
\newblock \emph{arXiv preprint arXiv:2306.12925}.

\bibitem[{Taal et~al.(2010)Taal, Hendriks, Heusdens, and Jensen}]{STOI}
Cees~H Taal, Richard~C Hendriks, Richard Heusdens, and Jesper Jensen. 2010.
\newblock A short-time objective intelligibility measure for time-frequency weighted noisy speech.
\newblock In \emph{2010 IEEE International Conference on Acoustics, Speech and Signal Processing}, pages 4214--4217. IEEE.

\bibitem[{Tay et~al.(2020)Tay, Dehghani, Bahri, and Metzler}]{Tay2020EfficientTA}
Yi~Tay, Mostafa Dehghani, Dara Bahri, and Donald Metzler. 2020.
\newblock \href {https://api.semanticscholar.org/CorpusID:221702858} {Efficient transformers: A survey}.
\newblock \emph{ACM Computing Surveys}, 55:1 -- 28.

\bibitem[{Vaswani(2017)}]{vaswani2017attention}
A~Vaswani. 2017.
\newblock Attention is all you need.
\newblock \emph{Advances in Neural Information Processing Systems}.

\bibitem[{Wang et~al.(2023{\natexlab{a}})Wang, Chen, Wu, Zhang, Zhou, Liu, Chen, Liu, Wang, Li, He, Zhao, and Wei}]{VALLE}
Chengyi Wang, Sanyuan Chen, Yu~Wu, Zi-Hua Zhang, Long Zhou, Shujie Liu, Zhuo Chen, Yanqing Liu, Huaming Wang, Jinyu Li, Lei He, Sheng Zhao, and Furu Wei. 2023{\natexlab{a}}.
\newblock \href {https://api.semanticscholar.org/CorpusID:255440307} {Neural codec language models are zero-shot text to speech synthesizers}.
\newblock \emph{ArXiv}, abs/2301.02111.

\bibitem[{Wang et~al.(2023{\natexlab{b}})Wang, Zhou, Zhang, Wu, Liu, Gaur, Chen, Li, and Wei}]{wang2023viola}
Tianrui Wang, Long Zhou, Ziqiang Zhang, Yu~Wu, Shujie Liu, Yashesh Gaur, Zhuo Chen, Jinyu Li, and Furu Wei. 2023{\natexlab{b}}.
\newblock Viola: Unified codec language models for speech recognition, synthesis, and translation.
\newblock \emph{arXiv preprint arXiv:2305.16107}.

\bibitem[{Wang et~al.(2024)Wang, Thakker, Chen, Kanda, Eskimez, Chen, Tang, Liu, Li, and Yoshioka}]{Speechx}
Xiaofei Wang, Manthan Thakker, Zhuo Chen, Naoyuki Kanda, Sefik~Emre Eskimez, Sanyuan Chen, Min Tang, Shujie Liu, Jinyu Li, and Takuya Yoshioka. 2024.
\newblock Speechx: Neural codec language model as a versatile speech transformer.
\newblock \emph{IEEE/ACM Transactions on Audio, Speech, and Language Processing}.

\bibitem[{Wu et~al.(2024)Wu, Chen, Lin, Chang, Chung, Liu, and Lee}]{tts_overview}
Haibin Wu, Xuanjun Chen, Yi-Cheng Lin, Kai-wei Chang, Ho-Lam Chung, Alexander~H Liu, and Hung-yi Lee. 2024.
\newblock Towards audio language modeling-an overview.
\newblock \emph{arXiv preprint arXiv:2402.13236}.

\bibitem[{Wu et~al.(2023)Wu, Gebru, Markovi{\'c}, and Richard}]{audiodec}
Yi-Chiao Wu, Israel~D Gebru, Dejan Markovi{\'c}, and Alexander Richard. 2023.
\newblock Audiodec: An open-source streaming high-fidelity neural audio codec.
\newblock In \emph{ICASSP 2023-2023 IEEE International Conference on Acoustics, Speech and Signal Processing (ICASSP)}, pages 1--5. IEEE.

\bibitem[{Yang et~al.(2023{\natexlab{a}})Yang, Liu, Huang, Tian, Weng, and Zou}]{yang2023hifi}
Dongchao Yang, Songxiang Liu, Rongjie Huang, Jinchuan Tian, Chao Weng, and Yuexian Zou. 2023{\natexlab{a}}.
\newblock Hifi-codec: Group-residual vector quantization for high fidelity audio codec.
\newblock \emph{arXiv preprint arXiv:2305.02765}.

\bibitem[{Yang et~al.(2023{\natexlab{b}})Yang, Tian, Tan, Huang, Liu, Chang, Shi, Zhao, Bian, Wu et~al.}]{Uniaudio}
Dongchao Yang, Jinchuan Tian, Xu~Tan, Rongjie Huang, Songxiang Liu, Xuankai Chang, Jiatong Shi, Sheng Zhao, Jiang Bian, Xixin Wu, et~al. 2023{\natexlab{b}}.
\newblock Uniaudio: An audio foundation model toward universal audio generation.
\newblock \emph{arXiv preprint arXiv:2310.00704}.

\bibitem[{Yang et~al.(2024)Yang, Jin, Tang, Han, Feng, Jiang, Zhong, Yin, and Hu}]{yang2024harnessing}
Jingfeng Yang, Hongye Jin, Ruixiang Tang, Xiaotian Han, Qizhang Feng, Haoming Jiang, Shaochen Zhong, Bing Yin, and Xia Hu. 2024.
\newblock Harnessing the power of llms in practice: A survey on chatgpt and beyond.
\newblock \emph{ACM Transactions on Knowledge Discovery from Data}, 18(6):1--32.

\bibitem[{Zhang et~al.(2024)Zhang, Yu, Li, Dong, Su, Chu, and Yu}]{mmllm}
Duzhen Zhang, Yahan Yu, Chenxing Li, Jiahua Dong, Dan Su, Chenhui Chu, and Dong Yu. 2024.
\newblock Mm-llms: Recent advances in multimodal large language models.
\newblock \emph{arXiv preprint arXiv:2401.13601}.

\bibitem[{Zhang et~al.(2023)Zhang, Zhang, Li, Zhou, and Qiu}]{speechtokenizer}
Xin Zhang, Dong Zhang, Shimin Li, Yaqian Zhou, and Xipeng Qiu. 2023.
\newblock Speechtokenizer: Unified speech tokenizer for speech large language models.
\newblock \emph{arXiv preprint arXiv:2308.16692}.

\bibitem[{Zhuang et~al.(2023)Zhuang, Liu, Pan, He, Weng, and Shen}]{zhuang2023survey}
Bohan Zhuang, Jing Liu, Zizheng Pan, Haoyu He, Yuetian Weng, and Chunhua Shen. 2023.
\newblock A survey on efficient training of transformers.
\newblock \emph{arXiv preprint arXiv:2302.01107}.

\end{thebibliography}

\appendix


\section{Experiments Illustration}

\subsection{Ablation Study}\label{app:ablation}

\begin{table}[h]
  \centering
  \begin{tabular}{ccc}
    \hline
    \textbf{Model $\backslash$  Metrics} & \textbf{WER $\downarrow$}  & \textbf{SIM $\uparrow$} \\
    \hline
    Ours       & \textbf{6.59}  &   \textbf{0.73}          \\
    $\,$ w/o SPT Joint Training       & 6.83  &   0.71         \\
    $\,$ w/o $\mathcal{L}_{spt}$ Pre-train   &  8.42  & 0.64           \\
    \hline
  \end{tabular}
  \caption{Ablation study of our model. }
  \label{tab:ablation}
\end{table}

We conduct the ablation study on our method. Table \ref{tab:ablation} shows the effect of continuous speech tokenizer pre-training, the impact of speech tokenizer joint training. Our method provides a complete and usable framework for continuous speech tokenizer and continuous speech token based TTS.

\subsection{Metrics}\label{app:metric}

In this section, we elucidate the details of the metrics used in this paper.


\textbf{WER} The Word Error Rate is caculated by performing ASR using Whisper Large \cite{Whisper} on the generated speech.

\textbf{SIM} The Speaker Similarity is obtained by first calculating the speaker embeddings of both the generated speech and the ground truth speech using WavLM-TDNN \cite{chen2022wavlm}, and then calculating the cosine similarity between the embeddings.

\textbf{EMoS} The Mean Opinion Score is a commonly used indicator for evaluating TTS quality. We use model-estimated EMoS \cite{EMOS} to provide a more general evaluation, and the open source library we utilize is located at \cite{SPEECHMETRICS}. 

\textbf{CLVP Score} The CLVP Score  measures the distance predicted by Contrastive Language-Voice Pretrained model between text and an audio clip where that text is spoken, and the open source library we utilize is located at \cite{TTS-scores} 

\textbf{STOI} The Short-Time Objective Intelligibility \cite{STOI} is used to reflect the clarity of speech, ranging from 0 to 1. The open source library we utilize is located at \cite{SPEECHMETRICS}.  

\textbf{Noisiness, Continuity, Loudness Quality, and Naturalness} The Noisiness, Continuity, Loudness, and Naturalness are calculated by NISQA \cite{NISQA} model, and the open source library we utilize is located at \cite{nisqa_github}

\section{Supplement to Related Work}

In this section, we briefly supplement the related work on speech tokenizers in the TTS field. Since the currently commonly used Encodec \cite{encodec} was proposed, many works have studied the form and properties of tokenizers such as Audiodec \cite{audiodec}. AudioLM \cite{audiolm} propose a hybrid tokenization scheme to achieve the combination of reconstruction quality and long-term structure. Hifi-codec \cite{yang2023hifi} propose the group-residual vector quantization (GRVQ) technique. SpeechTokenizer \cite{speechtokenizer} proposed to unify semantic and acoustic tokens by combining semantic distillation in the RVQ process. Concurrent work MELLE \cite{melle} proposed a TTS model based on mel-spectrogram feature extraction and vocoder. It is worth noting that MELLE also questioned the concept of discrete speech tokens. Unlike us, they removed the speech tokenizer. We retained this concept and conducted pre-training and ablation experiments on the continuous speech tokenizer.

\end{document}